# Evolution of magnetic and dielectric properties in Sr-substituted high temperature multiferroic YBaCuFeO$_5$

Surender lal[1], Sanjay K. Upadhyay[2], K. Mukherjee[1], C. S. Yadav[1]

[1]School of Basic Sciences, Indian Institute of Technology Mandi, Mandi 175005, Himachal Pradesh, India

[2]Tata Institute of Fundamental Research, Homi Bhabha Road, Colaba, Mumbai 400005, India

Email:kaustav@iitmandi.ac.in

## Abstract

We report the evolution of structural, magnetic and dielectric properties due to partial substitution of Ba by Sr in the high temperature multiferroic YBaCuFeO$_5$. This compound exhibits ferroelectric and antiferromagnetic transitions around 200 K and these two phenomena are presumed to be coupled with each other. Our studies on magnetic and dielectric properties of the YBa$_{1-x}$Sr$_x$CuFeO$_5$ (x = 0.0, 0.25 and 0.5) show that substitution of Sr shifts magnetic transition towards higher temperature whereas dielectric transition to lower temperature. These results points to the fact that magnetic and dielectric transitions get decoupled as a result of chemical pressure in form of Sr substitution. The nature of magnetodielectric coupling changes across the series with the presence of higher order coupling terms. Additionally in these compounds glassy dynamics of electric dipoles is observed at low temperatures.

**Introduction**

In last couple of decades, materials exhibiting multiferroic properties are being extensively investigated as these materials provide a platform to study the interesting physics associated with the mutual coupling between the spin and lattice degrees of freedom, and in view of their possible use in magnetoelectric device applications [1-9]. In general magnetism requires partially filled *d*- orbital and ferroelectricity requires empty *d*- orbital where polarization is observed due the loss of centre of symmetry. However, there can be number of possible ways by which the ferroelectricity can be induced which may be due to lone pair, site distortion, bond and magnetic ordering. Among multiferroics materials, some exhibit spontaneous appearance of electric polarization $P$ which is linked to the onset of incommensurate (ICM) antiferromagnetic order [3, 5, 7, 8]. However, such type of ordering is generally observed at low temperatures [9] thereby limiting the potential technological multifunctionalities of these materials.

Interestingly, Kundys *et al.* [10] reported electrical polarity induced by incommensurate antiferromagnetic order around 230 K in $YBaCuFeO_5$. These compounds along with CuO [6] are few rare examples where such a property is observed at high temperature. $YBaCuFeO_5$ belongs to the family of oxygen deficient double perovskite compounds which crystallizes in tetragonal structure [11]. The structure of the compound consists of $[CuFeO_5]_\infty$ double layers of the corner sharing $CuO_5$ and $FeO_5$ pyramids perpendicular to the *c*-direction and containing the $Ba^{2+}$ ions. These $[CuFeO_5]_\infty$ double layers are separated by $Y^{3+}$ planes. If the $Fe^{3+}$ and $Cu^{2+}$ ions are equally distributed among the pyramids the structure is centrosymmetric (Space group: P4/mmm). If the distribution of ions $Cu^{2+}/Fe^{3+}$ is asymmetric, a mirror plane containing the $Y^{3+}$ ions is lost and the structure becomes non-centrosymmetric (Space group: P4mm). The distortion in $CuO_5$ and $FeO_5$ pyramids alters the magnetic structure of $YBaCuFeO_5$. In this compound, commensurate antiferromagnetic (CM-AFM) ordering sets around 441 K ($T_{N1}$), which transforms to incommensurate antiferromagnetic (ICM-AFM) ordering below 230 K ($T_{N2}$) [10]. However, Kawamura *et al.* reported the ordering temperature in this compound to be at ~ 455 and 180 K [12], whereas Morin *et al.* reported these at ~ 440 and 200 K respectively [13]. Further, neutron diffraction studies have shown that the appearance of polarization below $T_{N2}$ is accompanied by replacement of the high-temperature collinear magnetic order by a circular inclined spinal order [13]. The observed properties of this compound are ascribed to the weak alternate ferromagnetic (FM) and antiferromagnetic (AFM) coupling along the *c*-direction. In this compound, external perturbation in the form of chemical pressure will change the structural parameters. This will directly tune the magnetic interactions, which will lead to interesting physical properties. Additionally, it may also bring the magnetic and dielectric transition temperatures closer to room temperature, which can be of immense technological importance.

With this aim, here we report the structural, magnetic and dielectric properties of YBa$_{1-x}$Sr$_x$CuFeO$_5$ (x =0.0, 0.25 and 0.5) compounds. Partial substitution of Sr in place of Ba leads to contraction of the unit cell, which develops positive chemical pressure. In our study, we found that YBaCuFeO$_5$ undergoes magnetic ordering $T_{N2}$ around 219 K while a dielectric anomaly was observed around 185 K. Interestingly, with an increase in Sr substitution, it was found that the magnetic transition ($T_{N2}$) shifts towards the higher temperature but the dielectric anomaly shifts to lower temperatures. This observation highlights the fact that with the Sr-substitution the magnetic transition and the dielectric anomaly get decoupled. The nature of magneto-dielectric coupling changes with Sr-substitution. Additionally these compounds show dipolar cluster glass type behavior in the low temperature region.

**Experimental**

Polycrystalline samples of YBa$_{1-x}$Sr$_x$CuFeO$_5$ (x = 0, 0.25, 0.5) were prepared by the solid state reaction method. The powders of the Y$_2$O$_3$, BaCO$_3$, SrCO$_3$ Fe$_2$O$_3$ and CuO were taken in stoichiometric ratio. The mixture was grounded and heated at 900 $^0$C for 36 hours. Samples were further re-grounded, pelletized and sintered at 1000 $^0$C. The structural analysis was carried out by x-ray diffraction (XRD, Cu K$\alpha$) using Rigaku Smart Lab instruments. The Rietveld refinement of the powder diffraction data was performed by FullProf Suite software [14].Temperature (*T*) (2−400 K) and magnetic field (up to *H* = 50 kOe) dependent magnetization (*M*) was measured with a Magnetic Property Measurements System (MPMS), Quantum design, USA. Complex dielectric permittivity was measured as a function of *T* (2−300 K) and *H* using Agilent E-4980A LCR meter with a homemade sample holder integrated with PPMS. Measurements were made on a thin parallel plate capacitor with Cu electrodes. Contacts were made using silver paints. The ac bias was 1 V and different frequencies (5−300 kHz) were used. The data were collected in the warming cycle (at the rate of 1 K min$^{-1}$).

**Results and discussion**

Rietveld refined x-ray diffraction patterns of the compounds are shown in the Fig. 1 and the refined parameters are tabulated in Table 1. It is observed that the volume of the unit cell decreases upon doping of the smaller size Sr atom, resulting in a positive chemical pressure. We have shown the schematic of the crystal structure in the Fig. 2, on the basis of low energy structure model mentioned in literature[13, 15] In this model, the super cell with $a_s = \sqrt{2} a_c$ and $c_s = 2c_c$ is shown with equal number of $CuO_5$ and $FeO_5$ pyramids with their respective positions. As seen from table 1, the respective separations between the $CuO_5$ and $FeO_5$ pyramids ($d_3$ and $d_4$) i.e. inter-bipyramidal distance increase upon Sr-substitution. It implies that the bypiramids moves further apart as the Sr concentration increases. Such structural distortion might play pivotal role in altering the magnetic coupling in the system.

Fig. 3 (a) shows the $T$ response of dc susceptibility ($\chi$) of the compounds in zero field cooled (ZFC) conditions measured at 100 Oe. To ascertain onset of the magnetic transition, we have plotted $d\chi/dT$ vs. $T$ curve as shown in the inset (i) of Fig. 3a. For $YBaCuFeO_5$, the CM-AFM to ICM-AFM magnetic transition $T$ is around 220 K. This transition has been observed in the range 180 - 230 K by various groups, as reported in literature [10, 12 and 13]. The $M$ ($H$) isotherms measured at 2 K (inset (ii) of Fig. 3 (a)), show linear behavior without hysteresis. Similar behavior was also observed at 300 K (not shown) which confirms the antiferromagnetic nature of the compound in the $T$ range of our measurement. It is observed that the transition shifts to higher $T$ ~ 264 K and 288 K (inset (i) of Fig 3 (a)) for $YBa_{0.75}Sr_{0.25}CuFeO_5$ and $YBa_{0.5}Sr_{0.5}CuFeO_5$ respectively (along with the linear behavior of $M$ ($H$) isotherms (inset (i) of Fig 3 (b)). Thus, it can be said that doping of Sr, distorts the commensurate nature of antiferromagnetic ordering and enhances CM to ICM AFM ordering $T$. To substantiate the above statement, we have to look into the structure of $YBaCuFeO_5$ [Fig. 3]. It is to be noted that this transition depends on the alignment of Cu, Fe ions in the bipyramids. As already reported [11], the Cu and Fe ions are aligned anti-parallelly along ab-plane in $YBaCuFeO_5$. Along c-direction, magnetic ions are aligned ferrmagnetically within bipyramids, whereas, the bypyramids are aligned anti-parallely. Thus, along c-direction the magnetic coupling is modified and on lowering $T$ gives rise to the formation of ICM-AFM structure. In the Sr doped systems, altered magnetic interaction due to variation in the inter-bipyramid distance and bipyramidal size, favors the stablization of IC-AFM state at higher $T$ in these compounds, in comparison to the parent compound.

To investigate the dielectric behavior of the compound series, we measured the $T$ response of complex dielectric permittivity in the range 2 to 300 K in zero magnetic field. The $T$ responses of the real part of dielectric permittivity ($\varepsilon'$) and the loss part (tan$\delta$) of the compound series, at 5 kHz are shown in the insets of Fig. 3 (b). For YBaCuFeO$_5$, dielectric anomaly was not observed in $\varepsilon'$ versus $T$ curve. Similar dielectric bahvior has been reported by Lai *et al.* on YBaCuFeO$_5$ single crystal [16]. However, the plot of d$\varepsilon'$/d$T$ versus $T$ (Fig 3 (b)) shows a weak anomaly around 185 K in our YBaCuFeO$_5$ compound, which is in agreement with literature [12]. Interestingly, for YBa$_{0.75}$Sr$_{0.25}$CuFeO$_5$ and YBa$_{0.5}$Sr$_{0.5}$CuFeO$_5$ the dielectric anomaly is observed at lower $T$ ~ 170 and 145 K for respectively. It is to be noted that the average size of bipyramid decreases while seperation between bipyramids increases upon Sr substituion, leading to the stablization of the spiral magnetic state [17]. It is possible that the modification in magnetic interactions upon Sr substitution reduces the electric polarization and dielectric anomaly shifts towards lower temperatures. However to validate this, $T$, and $H$ dependent microscopic studies are warranted.

The dielectric behavior of the compound series was investigated in the presence of a magnetic field. It was observed that under magnetic field dielectric behavior changes, which indicates the presence of magneto-dielectric effect (MDE). To study the MDE, the magnetic field response of $\varepsilon'$ was noted at selected $T$ in CM and ICM AFM regions. The results of such measurements at 5 kHz are shown in Fig. 4 (a –c), in the form of $\Delta\varepsilon'$ vs. $H$, where $\Delta\varepsilon' = [(\varepsilon'_H - \varepsilon'_{H=0})/\varepsilon'_{H=0}]$. The magnitude of $\Delta\varepsilon'$ for YBaCuFeO$_5$ compound is comparable to the value reported in the literature [10]. The observed behavior clearly reveals that there are qualitative changes in the behavior across $T$ and with Sr substitution. Further, the magnetic field response of $\Delta\varepsilon'$ suggests the presence of higher order MDE term in these compounds. The data were fitted using a combination of linear and quadratic term in $H$ by using the equation $MD \sim (\beta_1 H + \beta_2 H^2)$ [18, 19]. The coefficient of the quadratic term $\beta_2$ is plotted as a function of $T$ (insets of the Fig. 4 (a - c)). The value of the $\beta_2$ indicates the existence of nonlinear coupling along with linear coupling in the ICM-AFM region. Fig 4 (d) shows the $T$ response of $\Delta\varepsilon'$ at 90 kOe at 5 kHz. It was observed that $\Delta\varepsilon'$ is maximum near the vicinity of the dielectric peak in the ICM-AFM region for the compounds YBaCuFeO$_5$ and YBa$_{0.75}$Sr$_{0.25}$CuFeO$_5$. The MDE coupling reduces further as the $T$ is increased. For YBa$_{0.5}$Sr$_{0.5}$CuFeO$_5$ compound no such peak is observed and $\Delta\varepsilon'$ increases with increasing $T$. At 300 K and 90 kOe, it is observed that $\Delta\varepsilon'$ has negative values ~ 0.22% and 0.09% for YBaCuFeO$_5$ and YBa$_{0.75}$Sr$_{0.25}$CuFeO$_5$ respectively, and a positive value of ~

0.54% for $YBa_{0.5}Sr_{0.5}CuFeO_5$. Hence it can be said that the substitution of Sr changes the nature of the MDE coupling.

In addition to the above behaviors, a peak in $\varepsilon'$ was observed at ~ 33 K for 5 kHz in $YBaCuFeO_5$ compound. Similar behavior was also observed in Ref [10], but was not addressed. Interestingly, this peak $T$ increases as frequency is increased. Similar behavior was observed in $T$ response of $\varepsilon''$ of all the compounds. Also it is to be noted that the tanδ value in this region sufficiently small (≤ 0.07) indicating the intrinsic nature of this feature. To highlight this anomaly, $T$ response of $d\varepsilon''/dT$ measured at different frequencies for all the compounds is plotted in Fig. 5. The frequency dependence of the peak $T$ is found to follow the power law behavior of the form $\tau = \tau_0 (T/T_0 - 1)^{-zv}$ (Insets of Fig. 5), where $\tau_0$ is the characteristic relaxation time, $T_0$ is the glass transition $T$, and $zv$ is the exponent. Such a behavior is generally observed in compounds which exhibit the presence of glassy dynamics of electric dipoles. The values of $zv$ lie in the range of 4-5.5 and $\tau_0$ is of the order of $10^{-6}$ sec range, which are comparable to that for other dipolar glass compounds like $Fe_2TiO_5$, $BaTi_{1-x}Zr_xO_3$ etc. [18, 20].

## Summary

In summary, structural, magnetic, dielectric studies have been performed on a series of compounds $YBa_{1-x}Sr_xCuFeO_5$ (x =0.0, 0.25 and 0.5). Partial substitution of Sr in place of Ba results in the development of positive chemical pressure. As compared to the parent compound, the magnetic and dielectric transitions seem to be negatively coupled in the Sr substituted compounds, with the magnetic transition shifting to higher $T$ while the dielectric one shift to lower $T$. The nature of MDE coupling changes across the series and the behavior indicates the presence of higher order coupling terms. All the member of this series shows a glassy dynamics of electric dipoles in the low $T$ regions. This work gains importance considering the current interest in identifying multiferroics materials at high temperatures.

## Acknowledgement

The authors acknowledge Kartik K. Iyer, TIFR, for his help in measurements and experimental facilities of Advanced Materials Research Center (AMRC) IIT Mandi. SL acknowledges UGC, India for Fellowship.

**Table 1.** Crystal parameters for the compounds obtained from the rietveld refinement of X-ray diffraction data.

| Parameters | $YBa_{1-x}Sr_xCuFeO_5$ | | |
|---|---|---|---|
| | x = 0 | x = 0.25 | x = 0.5 |
| a(Å) | 3.871 | 3.860 | 3.850 |
| c(Å) | 7.662 | 7.659 | 7.651 |
| V(Å$^3$) | 114.83 | 114.14 | 113.45 |
| Size of $CuO_5/FeO_5$ bipyramid $d_1$ (Å) | 4.858 | 4.749 | 4.653 |
| Size of $FeO_5/CuO_5$ bipyramid $d_2$ (Å) | 4.857 | 4.933 | 4.996 |
| Separation between $CuO_5$ pyramids $d_3$ (Å) | 2.846 | 2.862 | 2.875 |
| Separation between $FeO_5$ pyramids $d_4$ (Å) | 2.766 | 2.775 | 2.779 |
| Bragg R-factor | 4.92 | 4.89 | 3.70 |
| RF-factor | 4.18 | 5.54 | 4.62 |
| $\chi^2$ | 2.18 | 1.42 | 1.49 |

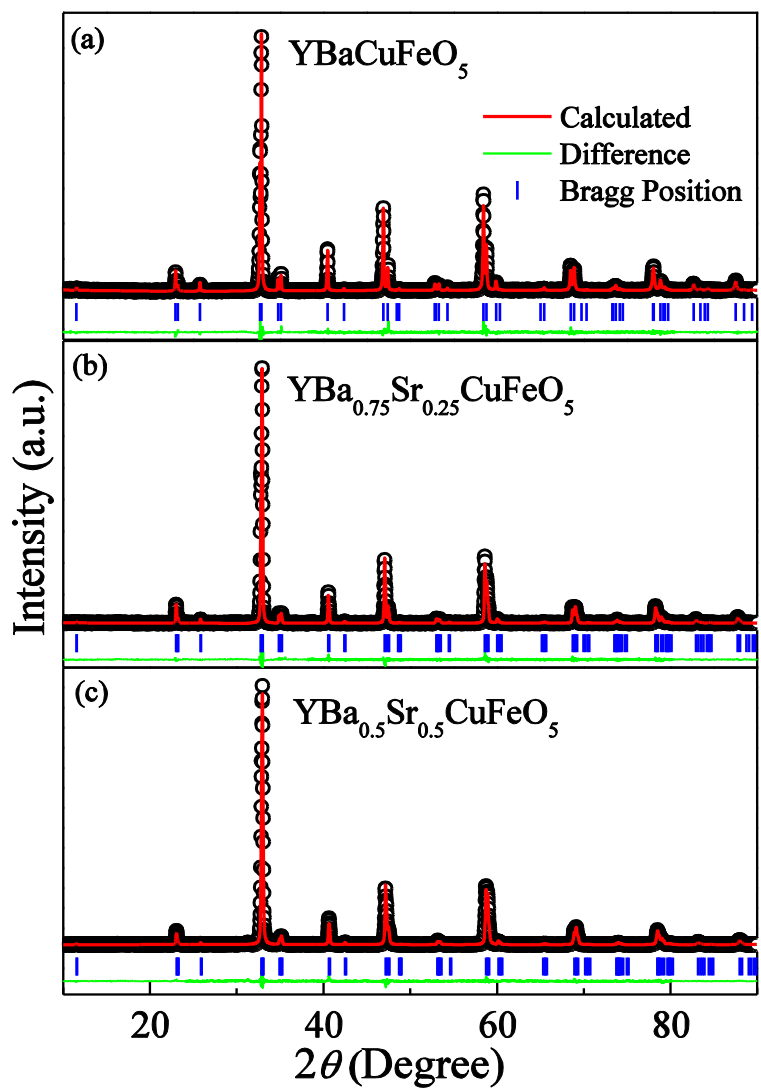

Fig. 1: Rietveld refined X-Ray diffraction patterns ($T = 300K$) of (a) $YBaCuFeO_5$, (b) $YBa_{0.75}Sr_{0.25}CuFeO_5$, and (c) $YBa_{0.5}Sr_{0.5}CuFeO_5$ compounds.

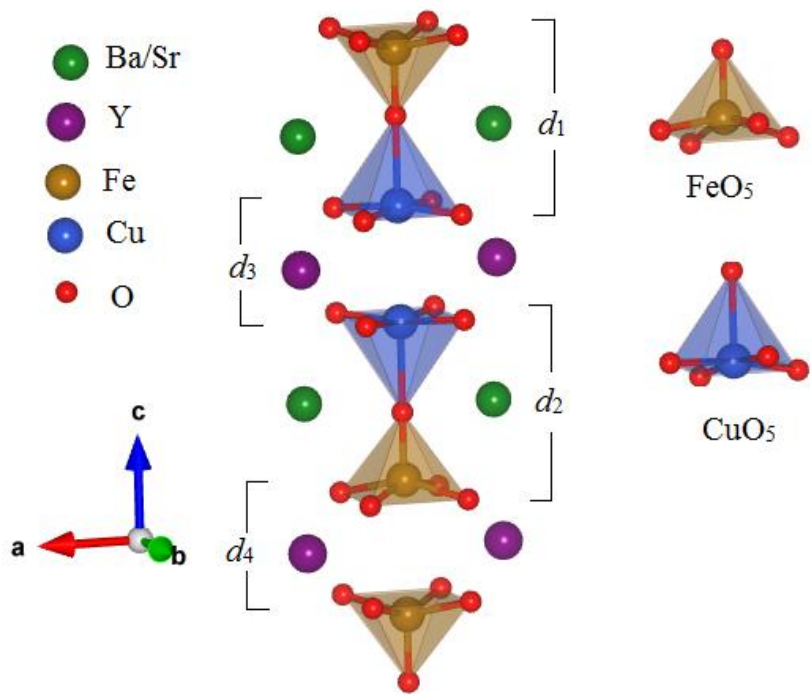

Fig 2: Schematic of the crystal structure showing the arrangement of atoms along with the $CuO_5$, and $FeO_5$ bipyramids.

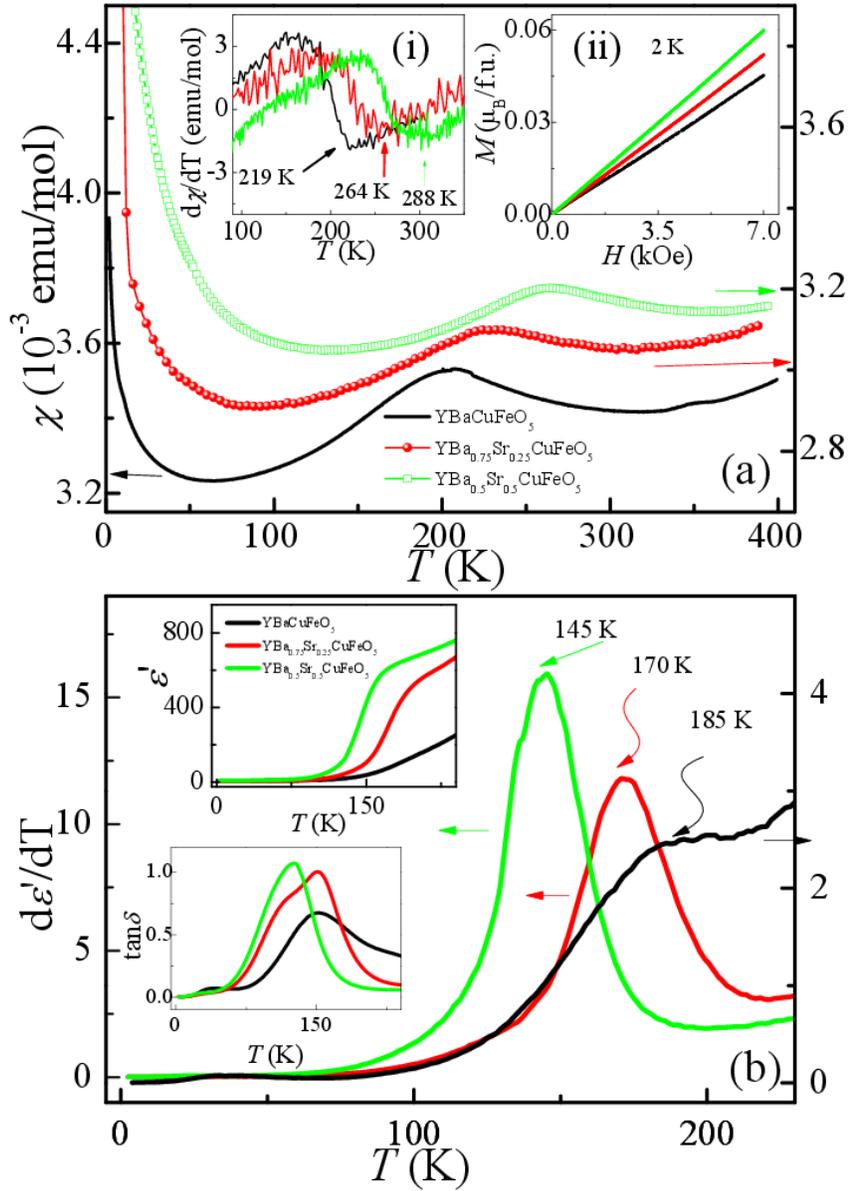

Fig. 3: (a) Temperature (*T*) dependent dc susceptibility ($\chi = M/H$) obtained under zero-field cooled condition at 100 Oe for YBa$_{1-x}$Sr$_x$CuFeO$_5$ (x = 0.0, 0.25 and 0.5) compounds. Insets (i) and (ii) show d$\chi$/d*T* vs. *T* and *M* vs. *H* at 2 K respectively. (b) *T* response of d$\varepsilon'$/d*T* at 5 kHz for all the compounds: Upper inset shows *T* response of dielectric constant ($\varepsilon'$). Lower inset shows the *T* dependence of tangent loss (tan$\delta$).

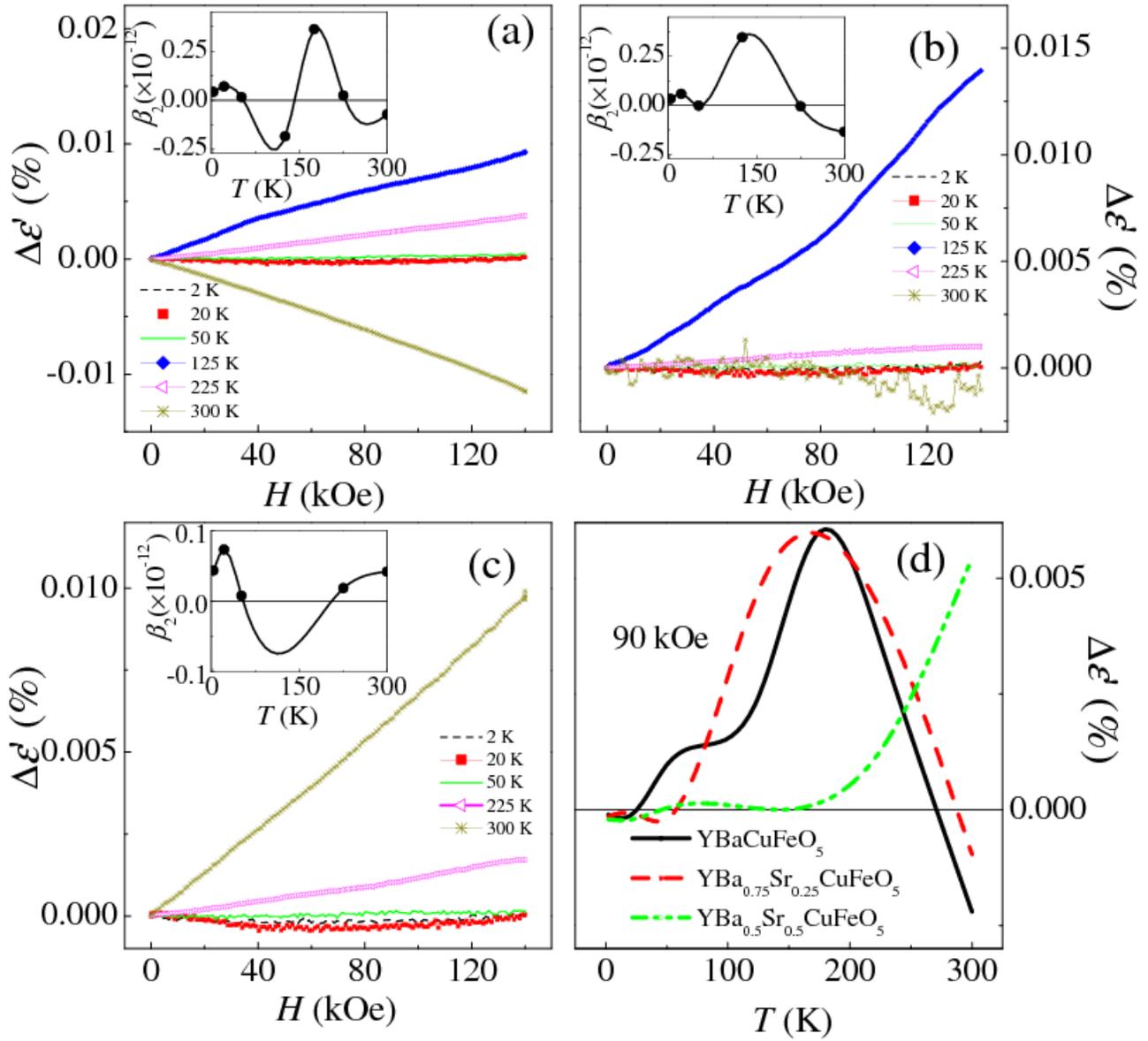

Fig. 4: (a)-(c) Relative change in dielectric constant (Δε′) under the application of magnetic field at different temperatures for YBa$_{1-x}$Sr$_x$CuFeO$_5$ (x = 0.0, 0.25 and 0.5) measured at 5 kHz. Inset: Coefficient of quadratic term ($\beta_2$) as function of temperature. (d) Variation of Δε′ with temperature for all the compounds at 90 kOe.

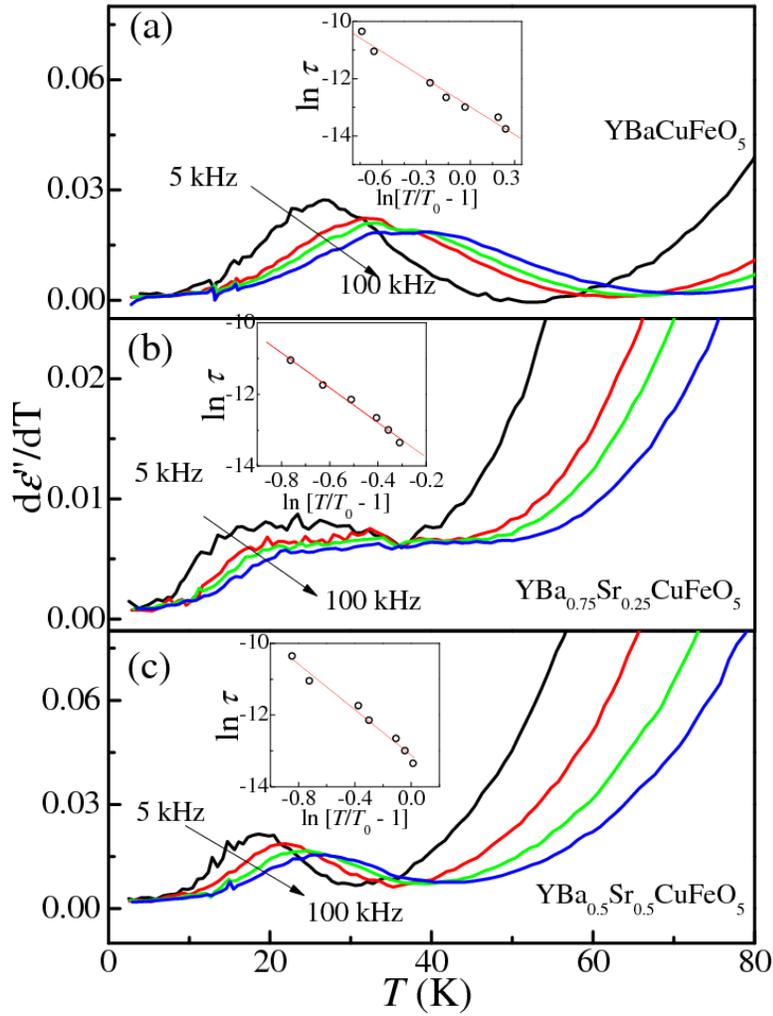

Fig. 5: (a)-(c) Temperature ($T$) dependence of the derivative of the imaginary part of dielectric permittivity $d\varepsilon''/dT$ at different frequencies 5 - 100 kHz for YBa$_{1-x}$Sr$_x$CuFeO$_5$ (x = 0.0, 0.25 and 0.5) in the $T$ range 2 - 80 K. Insets show the power law dependence ($\tau = \tau_0 \, (T/T_0 - 1)^{-zv}$) of the peak $T$ for the respective compounds.